\documentclass[conference]{IEEEtran}
\usepackage{graphicx, times, amsmath, amsfonts, amsthm, comment}
\usepackage{amssymb,epstopdf}
\usepackage{algorithmic}
\usepackage{algorithm}
\usepackage{color, soul}
\usepackage{mathtools}

\newcommand{\beq}{\begin{equation}}
\newcommand{\eeq}{\end{equation}}
\newcommand{\tbf}{\textbf}
\newcommand{\tit}{\textit}

\newcommand{\ud}{\mathrm{d}}

\newcommand*{\mathcolor}{}
\def\mathcolor#1#{\mathcoloraux{#1}}
\newcommand*{\mathcoloraux}[3]{%
  \protect\leavevmode
  \begingroup
    \color#1{#2}#3%
  \endgroup
}

\theoremstyle{plain}
\newtheorem{propcounter}{Proposition}
\newtheorem{proposition}[propcounter]{Proposition}
\theoremstyle{plain}
\newtheorem{lemcounter}{Lemma}
\newtheorem{lemma}[lemcounter]{Lemma}
\theoremstyle{plain}

\theoremstyle{plain}

\theoremstyle{plain}

\newcommand {\Ebb}{\mathbb{E}}

\newcommand {\Fcal}{\mathcal{F}}

\newcommand {\Scal}{\mathcal{S}}

\begin{document}

\title{Edge Caching for Cache Intensity under Probabilistic Delay Constraint }

\author{
\IEEEauthorblockN{Tachporn Sanguanpuak\IEEEauthorrefmark{1}, Sudarshan Guruacharya\IEEEauthorrefmark{2},
Nandana Rajatheva\IEEEauthorrefmark{1}, Matti Latva-aho\IEEEauthorrefmark{1}}
\IEEEauthorblockA{\IEEEauthorrefmark{1}Centre for Wireless Communications (CWC), University of Oulu, Finland; \\ \IEEEauthorrefmark{2}Dept. Elec. \& Comp. Eng., University of Manitoba, Canada}
\IEEEauthorblockA{Email: \{tachporn.sanguanpuak, nandana.rajatheva, matti.latva-aho\}@oulu.fi; sudarshan.guruacharya@umanitoba.ca}
}\maketitle

\begin{abstract}
In order to reduce the latency of data delivery, one of techniques is to cache the popular contents at the base stations (BSs) i.e. edge caching. However, the technique of caching at edge can only reduce the backhaul delay, other techniques such as BS densification will also need to be considered to reduce the fronthaul delay. In this work, we study the trade-offs between BS densification and cache size under delay constraint at a typical user (UE). For this, we use the downlink SINR coverage probability and throughput obtained based on stochastic geometrical analysis. The network deployment of BS and cache storage is introduced as a minimization problem of the product of the BS intensity and cache size which we refer to the product of ``\tit{cache intensity}'' under probabilistic delay constraint. We examine the cases when (i) either BS intensity or the cache size is held fixed, and (ii) when both BS intensity and the cache size are vary. For the case when both BS intensity and the cache size are variable, the problem become nonconvex and we convert into a geometric programing which we solve it analytically.

\end{abstract}


\section{Introduction} \label{section:introduction}
In the upcoming $5$G cellular networks, to achieve high throughput and very low latency as well as how to utilize the infrastructure more effectively are some of the main challenges in $5$G deployment. In~\cite{Cisco2016}, it was shown that more than $60\%$ percent of the traffic is due to the multiple downloads of popular video files. The redundancy of data transmission causes severe delay due to backhaul transmission, one of the methods to reduce backhaul delay is to employ the concept of proactive caching at the base stations (BSs). Proactive caching at edge has been introduced in~\cite{Wang_commag_2014}. Apart from the backhaul delay, the fronthaul delay also needs to be reduced in order to reduce the overall delay. As such, the technique of BS densification needs to be considered. By increasing the BS intensity, it can reduce the average number of user equipments (UEs) served by each BS, thus reducing the fronthaul delay.

In general, caching in wireless networks consists of two stages which are cache placement and data delivery. During the cache placement stage, by considering proactive caching, the most popular files will be cached at the BS during the off-peak traffic period. For the data delivery stage, the requested files from UEs will be delivered from cache storage at the BSs. In~\cite{Ejder_2015}, the benefits of popularity-based caching in terms of backhaul offloading was investigated. A collaborative hierarchical caching at both the cloud RAN and BSs with the aim of minimizing the delay of content delivery under UEs' quality-of-service (QoS) constraints was proposed in~\cite{Xinhua2017}. In~\cite{Zhiwen_2016}, various game theoretical approaches for wireless proactive caching was surveyed.
In \cite{CLaing2017}, the authors propose enhancing the QoE aware wireless edge caching with bandwidth provisioning in software-defined wireless networks. Caching as a service where
multiple service providers have to pay for the storages deployed at smallcell BSs that owned by MNOs, is studied in \cite{Hu_TWC2016}.
Auctions mechanism is proposed to solve this caching problems. The network virtualization, where the infrastructure providers (InPs) lease infrastructure and radio resources to the MNOs is provided in \cite{Tachporn_ICC2017}. The Cournot oligopoly market is used the model the multiple-seller multiple-buyer with infrastructure sharing deployment. However, the sharing of cache storage was not considered.

In this paper, we study the wireless aspects of caching at the cellular BS which owned by an MNO. The MNO aims to deliver the required quality-of-experience (QoE) to its UEs, in terms of latency of data delivery, regards to cache size and BS intensity. First, we obtain the downlink signal-to-interference-plus-noise-ratio (SINR) coverage probability and throughput based on  stochastic geometry then, using these results to analyze the end-to-end delay at UE. Second, assuming proactive caching placement, we formulate the minimization problem through the reduction of the BS intensity and/or cache storage with probablistic delay constraint at the typical UE of an MNO. We explore the cases, (i) when either BS intensity or the cache size is held fixed and (ii) when both BS intensity and cache size are vary. The optimization problem when both BS intensity and cache size are variable, becomes geometric programming and we can obtain solutions in closed form. The journal version of this paper is in \cite{Tachporn_TCOM2018}.

\section{System Model} \label{section:systemmodel}

Consider a set $\Phi_b$ of BSs owned by an MNO that are spatially distributed according to homogeneous Poisson point processes (PPPs). Each of the BSs is assumed to be equipped with a single antenna. Each BS is assumed to employ time division multiplexing access (TDMA) scheme. Thus, the BS serves a single UE in a given time slot. The maximum transmit power of each BS is $p_{\max}$. A UE subscribed to an MNO associates to the nearest BS. The net intensity of the BSs that a typical UE of the MNO can associate itself with is $\lambda$. The set of UEs $\Phi_u$ are assumed to be spatially distributed according to homogeneous PPPs with spatial intensity $\xi$ and let $\eta \in (0,1)$ be the activity level of a UE. The two point processes $\Phi_b$ and $\Phi_u$ are assumed to be independent of each other. Each UE is equipped with a single antenna. We will assume that the MNO has a bandwidth of $W$ Hz, which is divided into $L$ subchannels. Each BS operate in one of the $L$ available subchannels randomly assigned to it by the MNO. Thus,  the intensity of interfering BSs is given by $\lambda_I = \frac{\lambda}{L}$. 


\subsection{Caching Policy}
Finally, to enable edge caching, we assume that each BS can store $S$ number of files in its storage. For simplicity, we assume that all the files have equal size. Let $\mathcal{F} = \{f_1 \ldots, f_{F}\}$ be the set of files available for caching, where $F = |\mathcal{F}|$ is the total number of files. Based on the file popularity distribution and cache replacement policy at the edge, let $\Scal \subseteq \Fcal$ be the set of files cached at each BS, where $|\Scal| = S$ is the cache size. If a random file $f\in\Fcal$ is requested by a UE, then let the probability that the file-$f$ is available at the BS cache be $P_{\text{hit}}(S) = \Pr(f \in \Scal)$, which we refer to as ``\tit{hit probability}.''

We assume that the cache policy is to store the $S$ most popular files from $\Fcal$. We can model the popularity of the files by Zipf distribution given by
\beq
p_{d} = \frac{1/d^{\nu}}{\sum_{j=1}^{F} 1/j^{\nu}}.
\eeq
where $p_d$ is the probability of $d$-th most popular file being requested and the exponent $\nu>0$ reflects the skewness of the content popularity distribution. The larger of value $\nu$, the fewer popular contents hold a majority of the content requests.

The probability that the requested file $f \in \Fcal$ is stored in the cache is $P_{\text{hit}}(S) = \Pr(d \leq S)$, where $d$ is the popularity rank of the file $f$. We observe that $\Pr(d \leq S)$ is the cumulative distribution function (CDF) of the Zipf distribution. Hence, we can express $P_{\text{hit}}(S)$ as
\beq
P_{\text{hit}}(S) = \frac{\sum_{d=1}^S 1/d^{\nu}}{\sum_{j=1}^{F} 1/j^{\nu}} =  \frac{H_{S,\nu}}{H_{F,\nu}}.
\label{eqn:hit-probability}
\eeq
In (\ref{eqn:hit-probability}), we have concisely expressed $P_{\text{hit}}$ using generalized harmonic numbers, $H_{S,\nu}$ and $H_{F,\nu}$, where
\beq
H_{S,\nu} = \sum_{n=0}^{S-1} \frac{1}{(n+1)^\nu}.
\label{eqn:def-genearlized-harmonic-sum}
\eeq
The $H_{F,\nu}$ is defined similarly.

\subsection{Downlink SINR Coverage Probability and Goodput}
\label{sec:stogeoana}
Without loss of generality, we consider a typical UE of MNO located at the origin, which associates with the nearest BS. For convenience, let us label the nearest BS as BS-$0$. We assume that the message signal undergoes Rayleigh fading with the channel gain given by $g_0$. Furthermore, let $\alpha > 2$ denote the path-loss exponent for the path-loss model $r_0^{-\alpha}$, where $r_0$ is the distance between the typical UE and the nearest BS-$0$, $0 \in \Phi_b$. Finally, let $\sigma^2$ denote the noise variance; and $p$ denote the transmit power of all the BSs of the MNO. The downlink ${\rm SINR}$ at the typical UE is ${\rm SINR} = \frac{g_0 r_0^{-\alpha}p}{I + \sigma^2}.$

Since each BS employs TDMA scheme, the interference experienced by a typical UE associated with BS-$0$ comes from the transmit signal from other BSs to the UEs in the same time slot. Thus, $I = \sum_{j\in\Phi_b \backslash \{0\}} g_j r_j^{-\alpha}p$. Here $g_j$ is the channel gain between the typical UE and interfering BS-$j$, and $r_j$ is the distance between the typical UE and the interfering BS-$j$, where  $j\in\Phi_b \backslash \{0\}$. For a given threshold $T$, the SINR coverage probability for the typical UE is defined as
$P_c = \mathrm{Pr}({\rm SINR} > T).$
%

To find an analytical expression for $P_c$ under our system assumptions, we substitute $\lambda$ and $\lambda_I = \lambda/L$ in \cite[Prop. 1]{Tachporn_TMC2017}. We can express the coverage probability of a typical UE as,
\beq
P_c = \pi \lambda \int_0^{\infty}\exp \{-(Az + Bz^{\alpha /2})\} dz,
\label{eqn:coverage-integral}
\eeq
where the coefficients $A$ and $B$ are given by $A = \pi [\lambda_I(\beta -1) + \lambda]$ and $B = \frac{T \sigma^2}{p}$. Subsequently, we can evaluate (\ref{eqn:coverage-integral}) by using a simple closed form approximation, as given by \cite[Eqn. 4]{Sudarshan2016}, as $P_c \simeq  \pi \lambda \left[ A+ \frac{\alpha}{2} \frac{B^{2/\alpha}}{\Gamma\big(\frac{2}{\alpha}\big)} \right]^{-1}$. Therefore, we obtain
\begin{align}
P_c =& \left[ 1 + \frac{\beta - 1}{L} + \frac{\alpha}{2 \pi \lambda \Gamma\big(\frac{2}{\alpha}\big)} \left( \frac{T\sigma^2}{p} \right)^{2/\alpha} \right]^{-1},
\label{eqn:coverage-approx}
\end{align}
where $\Gamma(z)$ is Gamma function. For interference limited case, when $\sigma^2 \to 0$ or when $\lambda \to \infty$, (\ref{eqn:coverage-approx}) simplifies to
\beq
P_c \simeq \frac{L}{\beta + L - 1}
\label{eqn:coverage-approx-interferencelimited}
\eeq
Equation (\ref{eqn:coverage-approx-interferencelimited}) is independent of $\lambda$. Also, as $L \to \infty$, we have from (\ref{eqn:coverage-approx-interferencelimited}) that $P_c \to 1$. Next, we introduce the performance metric in terms of throughput. The throughput of the typical UE served by the nearest  BS is
\beq
G = \frac{P_c W}{L} \log_2 (1+T),
\label{eqn:goodput}
\eeq
where $P_c$ is the downlink coverage probability, $W$ is bandwidth. We can approximate the throughput (\ref{eqn:goodput}) using  (\ref{eqn:coverage-approx}) in general, or  (\ref{eqn:coverage-approx-interferencelimited}) for the interference limited case.


\section{Delay Modeling}
\label{subsec:Proactive_Cache}

\subsection{Expected Fronthaul Delay}
\label{subsec:fronthaul_delay}

The delay in the transmission of a file between BS and UE is referred to \tit{fronthaul delay}. If a file requested by a UE is available in cache of the serving BS, then the delay incurred during transmission of the file is only due to the fronthaul. This delay is contributed by a number of factors, including finite channel capacity, transmission success rate, size of the file, and the number of UEs in a cell.
The potential delay due to the channel is the reciprocal of the goodput, $1/G$, in seconds per bit. If there are $N$ UEs in the cell being served simultaneously, since the BS deploys TDMA scheme, the goodput per UE is $G/N$. Thus, the potential delay for a UE is $N/G$. Hence, in order to transfer a file of fixed size $x_f$, the total fronthaul delay, is $D_{\text{fh}} = \frac{N x_f}{G}$.

Here, only $N$ is the random variable. The expected number of UEs inside an average Voronoi cell formed by the PPP of the BS, $\Phi_b$, is given by $\frac{\xi}{\lambda}$ \cite[Eqn 20]{Arvanitakis2015}. Thus, the number of UEs being served is
\beq
\Ebb[N] = \frac{\eta \xi}{\lambda},
\label{eqn:avgUE}
\eeq
where $\xi$ is the intensity of the UEs and $\eta \in (0,1)$ is the probability that an UE will request service from the BS. Normally, $\eta$ is quite small, allowing the MNO to retain a large number of idle UEs to be potentially served by a single BS by its limited infrastructure.  Hence, the average fronthaul delay is given by
\beq
\Ebb[D_{\text{fh}}]  = \frac{\Ebb[N] x_f}{G}  = \frac{\eta \xi x_f}{\lambda G}.
\label{eqn:fronthaul-delay}
\eeq

Equation (\ref{eqn:fronthaul-delay}) confirms our intuitive understanding that BS densification leads to lower fronthaul delay.

\subsection{Expected Backhaul Delay}
\label{subsec:backhaul_delay}
When the requested file from the UE is \tit{not} available in cache of the serving BS, then the file needs to be fetched from cloud server to the BS. Let us assume that the BSs connect to cloud via backhaul (fiber optic, ethernet, and T1). We can model the process of a BS fetching the contents from the cloud server as a $G/G/m$ queue. Let $\tau$ be the mean service time at a single server. That is, the average time taken for the server to deliver $x_f$ bits of information to the BS. The expected backhaul delay can be given as \cite[Eqn 2.14]{Whitt1993},
\beq
\Ebb[D_{\text{bh}}] \approx \left( \frac{c_a^2 + c_s^2}{2} \right) \Ebb[W(M/M/m)] + \tau
\label{eqn:backhaul-delay}
\eeq
where $\Ebb[W(M/M/m)] \approx \tau (\rho^{\sqrt{2(m+1)}-1})/(m(1-\rho))$ is the expected waiting time of $M/M/m$ queue. Here, $\phi$ denotes the mean arrival rate of file requests to the server, $\mu = m/\tau$ is service rate of the server, and $\rho = \phi/(m \mu)$ is the server utilization. Also, $c_a$ and $c_s$ are coefficient of variations of the inter-arrival time and the service time, respectively. For the stability of the cloud server queue, the condition $\rho < 1$ must be satisfied. When $m=1$, the  above approximation yields $\Ebb[D_{\text{bh}}] \approx  \left( \frac{c_a^2 + c_s^2}{2} \right) \left( \frac{\rho}{1-\rho} \right) \tau + \tau$.


\subsection{Expected Total Delay}
\label{subsec:Total_delay}
The delay experienced by a UE while downloading a file is only due to the fronthaul, $D_{\text{fh}}$, if the requested file is already cached at its serving BS. If this is not the case, then the delay experienced by the UE is the sum of the fronthaul and backhaul delay, $D_{\text{fh}} + D_{\text{bh}}$. Since the availability of a file in the BS cache is given by the hit probability, $P_{\text{hit}}$, the expected total delay is given by the law of total expectation as
 \begin{align}
\Ebb[D] &=  \Ebb[D_{\text{fh}}] + \Ebb[D_{\text{bh}}] (1 - P_{\text{hit}}(S)) \label{eqn:expected-total-delay}
 \end{align}

Since $P_{\text{hit}}$ depends on the cache size $S$, this implies that the minimum expected delay that we can achieve by changing only the cache size is $\Ebb[D_{\text{fh}}]$. This bound also suggests that we cannot impose an arbitrarily lower bound on the total delay by only adding cache to the BSs, which allows us to eliminate the backhaul delay only. In terms of latency, the fronthaul presents the ultimate bottleneck for cache based cellular systems. Therefore, we also need to decrease the number of BSs deployed to lower the fronthaul delay. Thus, we have cache versus base station deployment scenario.

\section{Problem Formulation}
\label{subsec:problem_formulation}
In this section, we consider the trade-offs between cache storage $S$ and the BS intensity $\lambda$. We will refer to the product $\lambda S$ as ``\tit{cache intensity}''. This is the amount of cache per unit area of the MNO. It allows us to capture the tradeoff between adding more BSs versus adding more cache. We will formulate an optimization problem so as to minimize the cache intensity while satisfying the latency constraint for a typical UE, while assuming that the size of the cache is large. The optimization problem is as follows:
\begin{align}
\label{eqn:min_lambda_S}
\text{min}_{\lambda, S}  & \quad  \lambda S \\
\text{s.t}. & \quad \text{Pr}(D  \geq D_{\text{th}}) \leq \gamma,
\label{eqn:geometric_prog} \\
& \quad S \leq F \label{eqn:cache-constraint}
\end{align}
where $\lambda \geq 0$ and $S \geq 0$. Here,  (\ref{eqn:geometric_prog}) is a probabilistic constraint that limits the latency above some threshold value $D_{\text{th}}$ to probability $\gamma \in (0,1)$. To make the problem more tractable, we have from Markov's inequality
\begin{align}
\text{Pr}(D  \geq  D_{\text{th}}) \leq \frac{\Ebb[D]}{ D_{\text{th}}}.
\label{eqn:Markov_in_Equal}
\end{align}

Using the Markov's inequality (\ref{eqn:Markov_in_Equal}), we can linearize the probabilistic constraint in (\ref{eqn:geometric_prog}) as
\beq
\Ebb[D] \leq \gamma D_{\text{th}}.
\label{eqn:tot_cache_Th}
\eeq
Substituting the expression for $\Ebb[D]$ from (\ref{eqn:expected-total-delay}) into  (\ref{eqn:tot_cache_Th}), we obtain after some algebra
\begin{align}
 1 -  \frac{\gamma D_{\text{th}} - \Ebb[D_{\text{fh}}] }{\Ebb[D_{\text{bh}}]} &\leq P_{\text{hit}}(S).
\label{eqn:expected_cache2}
\end{align}

Thus, in (\ref{eqn:expected_cache2}), we have succeeded in modifying the statement about the delay constraint into an equivalent statement concerning the cache size. Since $P_{\text{hit}}(S) \leq 1$, the left-hand-side of (\ref{eqn:expected_cache2}) is less than the unity, we can express (\ref{eqn:expected_cache2}) as,
\beq
\gamma D_{\text{th}} \geq \mathbb{E}[D_{\text{fh}}]
\label{eqn:sufficient-condition}
\eeq
This leads to a fundamental lemma about cache based cellular system:

\begin{lemma}
If $\mathbb{E}[D_{\text{fh}}] \leq \gamma D_{\text{th}}$, then the constraint (\ref{eqn:geometric_prog}) is feasible for some $S$ such that $S \leq F$.
\label{lemma:feasibility-of-delay-constraint}
\end{lemma}


\tbf{Lemma \ref{lemma:feasibility-of-delay-constraint}} gives us the sufficient condition under in which both the constraints (\ref{eqn:geometric_prog}) and (\ref{eqn:cache-constraint}) can be feasible. Although the delay constraint (\ref{eqn:tot_cache_Th}) is over the total delay, we see that the fronthaul delay plays the most crucial part.


In equation (\ref{eqn:sufficient-condition}), since $\mathbb{E}[D_{\text{fh}}]$ is a constant for given $\lambda$, we see that $\gamma$ and $D_{\text{th}}$ should be at least inversely proportional to each other.  If we have $\mathbb{E}[D_{\text{fh}}] = \gamma D_{\text{th}}$, then the cache size is $S = F$. That is, the BS should cache all the available files in $\Fcal$. This is an unrealistic expectation in practice. Hence, realistically, it should be the case that $\mathbb{E}[D_{\text{fh}}] < \gamma D_{\text{th}}$ so that $S < F$.

If we substitute the expression for $\mathbb{E}[D_{\text{fh}}]$ from  (\ref{eqn:fronthaul-delay}) in (\ref{eqn:sufficient-condition}), we obtain a minimum bound for the BS intensity required for the feasibility of (\ref{eqn:sufficient-condition}), and hence (\ref{eqn:geometric_prog}), as
\beq
\lambda \geq \frac{\eta \xi x_f}{\gamma D_{\text{th}} G}.
\label{eqn:lambda-lower-bound}
\eeq
This gives us a relationship between the intensity of the BS $\lambda$ and the UE $\xi$ for the delay constraint to be feasible for some $S$ such that $S<F$.




To facilitate further analysis, we first give the following lemma on the asymptotic approximation for the hit probability:

\begin{lemma}
When the size of cache $S$ is large and $\nu \neq 1$, the probability that the requested file $f\in \Fcal$ is in cache is asymptotically given by
\beq
P_{\text{hit}}(S)  \sim  \frac{1}{H_{F,\nu}} \left[ \zeta (\nu) - \frac{(S+1)^{1-\nu}}{\nu -1}\right],
\label{eqn:Phit_asymptotic}
\eeq
where $\zeta(\nu)$ is the Riemann zeta function.
\label{lemma:hit-prob}
\end{lemma}

\begin{IEEEproof}
See \tbf{Appendix}.
\end{IEEEproof}

\begin{figure}[t]
\centering
\includegraphics[height=2.4 in, width=2.6 in, keepaspectratio = true]{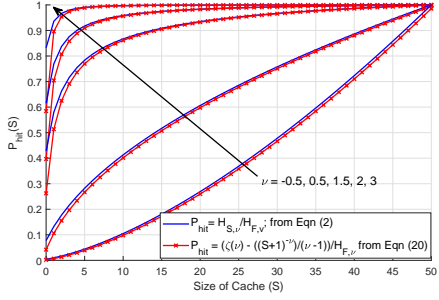}
\caption{Hit probability versus size of cache ($S$).}
\label{fig:Hit_prob:S}
\end{figure}

Note that although it was stipulated that $\Re(\nu) > 1$ during the definition of the Hurwitz zeta function in (\ref{eqn:hurwitz-zeta-fun}), the Riemann zeta function $\zeta(\nu)$ has a unique analytic continuation to the entire complex plane, excluding $\nu = 1$, which corresponds to a simple pole \cite{Sondow2017a}. Thus, so long as $\nu \neq 1$, the formula (\ref{eqn:Phit_asymptotic}) is applicable for any Zipf's exponent $\nu > 0$. In Fig. \ref{fig:Hit_prob:S}, we compare hit probability in (\ref{eqn:hit-probability}) using the exact value of $fH_{s,\nu}$ from (\ref{eqn:hurwitz-zeta-harmonic-sum-2}) and the asymptotic approximation for hit probability in (\ref{eqn:Phit_asymptotic}). It can be seen that when the size of cache is big, the asymptotic approximation holds tight with the exact one.

\section{Solutions and Tradeoff Analysis}
\label{sec:tradeoff-analysis}

Since we have a closed form approximation for $P_{\text{hit}}(S)$, as given by \tbf{Lemma \ref{lemma:hit-prob}}, we can now proceed to analyze the tradeoffs between cache size and BS intensity. We will first examine the case when either one of  $\lambda$ or $S$ is held fixed; after which, we will examine when both $\lambda$ and $S$ can vary.

\subsection{When either $\lambda$ or $S$ is fixed.}

Here we consider a modification of problem (\ref{eqn:min_lambda_S}), where instead of optimizing with respect to both $\lambda$ and $S$, we hold one of the term fixed. In \tbf{Proposition \ref{prop:optimal-cache}}, we will assume $\lambda$ to be fixed and performed the optimization with respect to $S$. In \tbf{Proposition \ref{prop:optimal-lambda}}, we will consider the case when $S$ is held fixed and $\lambda$ is optimized. Since left-hand-side term in (\ref{eqn:expected_cache2}) is a constant, let us denote it by $C =1 -  \frac{\gamma D_{\text{th}} - \Ebb[D_{\text{fh}}] }{\Ebb[D_{\text{bh}}]}$.


\begin{proposition}
For a fixed value of $\lambda$ such that (\ref{eqn:lambda-lower-bound}) is satisfied and that $\nu \neq 1$, the required cache size $S^{*}$ is at most
\beq
S^{*} = \left[(\nu -1)(\zeta (\nu) - C H_{F,\nu})\right]^{\frac{1}{1-\nu}} -1
\label{eqn:optcache_fixedlambda}
\eeq
\label{prop:optimal-cache}
\end{proposition}

\begin{IEEEproof}
By substituting the expression for $P_{\text{hit}}(S)$ from (\ref{eqn:Phit_asymptotic}) in \textbf{Lemma \ref{lemma:hit-prob}} into  (\ref{eqn:expected_cache2}), we can solve for $S$ to obtain the desired result.
\end{IEEEproof}


\begin{proposition}
For given fixed size of cache, $S$, such that $\nu \neq 1$ and $P_{\text{hit}}(S) > 1 - \frac{\gamma D_{\text{th}}}{\Ebb[D_{\text{bh}}]}$, the required BS intensity $\lambda^{*}$ is at least,
\begin{align}
\lambda^* = \frac{\eta \xi x_f}{G [\gamma D_{\text{th}} - \Ebb[D_{\text{bh}}]( 1- P_{\text{hit}}(S))] }
\label{eqn:optlambda_fixedS}
\end{align}
\label{prop:optimal-lambda}
\end{proposition}

\begin{IEEEproof}
Substituting the expression for the expected fronthaul delay (\ref{eqn:fronthaul-delay}) into (\ref{eqn:expected_cache2}) and solving for $\lambda$, we obtain the desired result. For the positivity of $\lambda$, the denominator of (\ref{eqn:optlambda_fixedS}) must be greater than zero,  $\gamma D_{\text{th}} - \Ebb[D_{\text{bh}}]( 1- P_{\text{hit}}(S)) > 0$. Re-arranging the terms gives us the sufficient condition $P_{\text{hit}}(S) > 1 - \frac{\gamma D_{\text{th}}}{\Ebb[D_{\text{bh}}]}$.
\end{IEEEproof}

%

\subsection{When both $\lambda$ and $S$ are variable}
We will now solve the general case when both $\lambda$ and $S$ is jointly optimized. For large $S$, we can recognize the optimization problem (\ref{eqn:min_lambda_S}) -- (\ref{eqn:geometric_prog}) as a geometric programming problem \cite{Richard1967}. In the following, we will first express the primal problem in (\ref{eqn:geometric_prog}) in the standard form a geometric program; after which, we will give the solution to the problem via its dual problem.

First, we expand $\mathbb{E}[D]$ in (\ref{eqn:expected-total-delay}) using (\ref{eqn:fronthaul-delay}) and (\ref{eqn:Phit_asymptotic}) in terms of $\lambda$ and $S$ as
\begin{align}
 \Ebb[D] &=   \frac{\eta \xi x_f}{G \lambda}  + \Ebb[D_{\text{bh}}] \left[ 1-\frac{1}{H_{F,\nu}} \left( \zeta (\nu) - \frac{(S+1)^{1-\nu}}{\nu -1}\right) \right] \nonumber\\
    &= C_1 + \frac{C_2}{\lambda} + C_3 (S+1)^{1-\nu}
\label{eqn:totcache_withC}
\end{align}
where $C_1 = \Ebb[D_{\text{bh}}]\left( 1 - \frac{\zeta(\nu)}{H_{F,\nu}}\right)$, $C_2 = \frac{\eta \xi x_f}{G}$, and $C_3 = \frac{ \mathbb{E}[D_{\text{bh}}]}{(\nu -1) H_{F,\nu}}$. Substituting (\ref{eqn:totcache_withC}) in the constraint (\ref{eqn:tot_cache_Th}), $\Ebb[D] \leq \gamma D_{\text{th}}$, we obtain,
\begin{align}
 & C_1 + \frac{C_2}{\lambda} + C_3 (S+1)^{1-\nu} \leq \gamma D_{\text{th}} \nonumber\\
\mathrm{or,}\; & \left( \frac{C_2}{\gamma D_{\text{th}} - C_1} \right) \frac{1}{\lambda} + \left(\frac{C_3}{\gamma D_{\text{th}} - C_1}\right)(S+1)^{1-\nu} \leq 1 \nonumber \\
\therefore \: & Q \lambda^{-1} + V t^{1-\nu} \leq 1
\label{eqn:geometric_constraint}
\end{align}
where $Q = \frac{C_2}{\gamma D_{\text{th}} - C_1}$, $V= \frac{C_3}{\gamma D_{\text{th}} - C_1}$ and $t= S+1$.

Since we consider the case when the storage size $S$ is large, we can approximate $t \approx S$. Thus, substituting the transform variable $t$ in the objective function (\ref{eqn:min_lambda_S}), substituting (\ref{eqn:geometric_constraint}) in the constraint (\ref{eqn:tot_cache_Th}), and substituting (\ref{eqn:lambda-lower-bound}) in the constraint (\ref{eqn:cache-constraint}), we can express the primal problem (\ref{eqn:geometric_prog}) as a geometric program:

\begin{proposition}
Assuming $S$ to be large and that $Q > 0$, $V > 0$ and $\nu \neq 1$, we can transform the problem (\ref{eqn:min_lambda_S}) -- (\ref{eqn:cache-constraint}) into an equivalent geometric programming problem
\begin{align}
\label{eqn:min_lambda_t}
\text{min}_{\lambda, t}  & \quad  g = \lambda t \\
\text{s.t}. & \quad Q \lambda^{-1} + V t^{1-\nu} \leq 1,  \\
& \quad R \lambda^{-1} \leq 1,
\end{align}
where $R = \frac{C_2}{\gamma D_{\text{th}}}$.
\end{proposition}

In geometric programming, when the orthogonality and normality conditions with dual variables $\delta_i$ are satisfied, the maximum of dual function is equal to the minimum of primal function $g$ \cite{Richard1967}. As such, we can express the dual maximization problem as,
\begin{align}
\label{eqn:dual_function}
\text{max}_\delta & \: q  = \bigg(\frac{1}{\delta_1}\bigg)^{\delta_1} \bigg(\frac{Q}{\delta_2}\bigg)^{\delta_2} \bigg(\frac{V}{\delta_3}\bigg)^{\delta_3} \bigg(\frac{R}{\delta_4}\bigg)^{\delta_4} (\delta_2 + \delta_3)^{\delta_2 + \delta_3} (\delta_4)^{\delta_4}\\
\text{s.t}. &\quad \delta_1 = 1 \label{eqn:geo_normality} \\
               &  \begin{pmatrix}
                 1 & -1 & 0 & -1\\
                  1 & 0 & 1 - \nu & 0
                \end{pmatrix}
                \begin{pmatrix}
                \delta_1 \\ \delta_2 \\ \delta_3 \\ \delta_4
		   \end{pmatrix}
            = 0,
\label{eqn:geo_orthogonality}
\end{align}
where $\delta_i \geq 0$ for $i = 1, \ldots, 4$. The degree of difficulty of this geometric program is 1. In our case, (\ref{eqn:geo_normality}) gives the normality condition while (\ref{eqn:geo_orthogonality}) gives the orthogonality condition. In geometric programming, we are focused on finding the optimal point of the dual variables $\boldsymbol{\delta}^{*} = (\delta_1^{*}, \delta_2^{*}, \delta_3^{*}, \delta_4^*)$ that maximizes the dual function $q$ subject to the orthogonality and normality conditions. Note that this dual problem is convex program with a concave objective function and linear constraints.

Using (\ref{eqn:geo_normality}) and (\ref{eqn:geo_orthogonality}), we can directly solve for $\boldsymbol{\delta}^{*}$. Here, matrix multiplication from (\ref{eqn:geo_orthogonality}) yields $\delta_1 - \delta_2 - \delta_4 = 0$, and  $\delta_1 - (1-\nu)\delta_3 =0$. Since $\delta_1 = 1$, we have $\delta_3 = \frac{1}{\nu - 1}$ and $\delta_2 + \delta_4 = 1$. Let $\delta_2 = r$, so that $\delta_4 = 1- r$. Since $\delta_2 \geq 0$ and $\delta_4 \geq 0$, we then have a bound over $r$ as $0 \leq r \leq 1$.

Substituting the values of $\delta$'s in the dual problem, we obtain a simpler problem constrained over a single variable $r$ as
\begin{align}
\label{eqn:modified-dual-problem}
\text{max}_r & \: q =  \bigg(\frac{Q}{r}\bigg)^{r} \bigg(\frac{V}{\nu - 1}\bigg)^{\nu - 1} \bigg(r + \frac{1}{\nu-1}\bigg)^{r + \frac{1}{\nu-1}} (1-r)^{1-r} \\
\text{s.t}. & \quad 0 \leq r \leq 1.
\label{eqn:bound-on-r}
\end{align}

Finding an analytical solution to the above problem is not possible. Hence, we must resort to numerical approaches. For the optimal primal variables $\lambda^*$ and $t^*$, we have
\begin{align*}
\lambda^* t^* &= \delta_1^* q^* = q^* \\
Q (\lambda^*)^{-1} &= \delta_2^* q^* = r^* q^*\\
V (t^{*})^{1-\nu} &= \delta_3^* q^* = \frac{q^{*}}{\nu -1} \\
R (\lambda^*)^{-1} &= \delta_4^* q^* = (1-r^*) q^*
\end{align*}
Adding the second and fourth row we have $Q (\lambda^*)^{-1} + R (\lambda^*)^{-1} = q^*$, which we can solve to obtain $\lambda^* = \frac{Q+R}{q^*}$. Also, we have $t^* = \left(\frac{V(\nu-1)}{q^*}\right)^{1/(\nu-1)}$. Note that for $t^*$ to be positive, we must have $\nu > 1$. Hence we have the following proposition:

\begin{proposition}
Assuming $Q > 0$, $V>0$ and $\nu > 1$, the optimal solution to (\ref{eqn:min_lambda_S}) -- (\ref{eqn:cache-constraint}) is $\lambda^* = \frac{Q+R}{q^*}$ and $S^* = \left(\frac{V(\nu-1)}{q^*}\right)^{1/(\nu-1)}$, where $q^*$ is the optima of the one dimensional problem (\ref{eqn:modified-dual-problem}) -- (\ref{eqn:bound-on-r}).
\label{prop:optSoptlambda}
\end{proposition}


Nonetheless, if (\ref{eqn:modified-dual-problem}) is monotonically increasing in $0 \leq r \leq 1$, then the maxima is at the boundary $r = 1$, and the optimal value at $r=1$ will be
\[ q^* = Q \left(\frac{V}{\nu - 1}\right)^{\nu - 1} \left(\frac{\nu}{\nu-1}\right)^{\frac{\nu}{\nu-1}}. \]
For the monotonicity of $q$ within $r\in[0,1]$, it is sufficient to check if $q'(r) >0$ at $r=1$. 
Since $Q, R, \nu$ are all positive, the derivative of $\log(q)$ with respect to $r$ is
\[ \frac{\ud \log(q)}{\ud r}  = \log\left(\frac{Q}{R}\right) + \log\left(1+ \frac{1}{r(v - 1)}\right). \]
Evaluating at $r=1$, if we have
\[ \left. \frac{\ud \log(q)}{\ud r} \right|_{r=1} = \log \left(\frac{\nu Q}{R(\nu-1)}\right) > 0, \]
then we can conclude that $r^* =1$.

\section{Numerical Results}
\label{subsec:results}
In this section, we evaluate the optimal size of cache ($S^*$) from \tbf{Proposition \ref{prop:optimal-cache}}, the optimal BS intensity ($\lambda^*$) from \tbf{Proposition \ref{prop:optimal-lambda}} and both of $S^*$ and $\lambda^*$ from \tbf{Proposition \ref{prop:optSoptlambda}}. We investigate the trade-offs between the size of cache, BS intensity and the latency. The baseline setting of simulation environments is as follows : the transmit power of BS is $p = 10 \text{dBm}$, $\sigma^2 = -150$ dBm, user intensity is $\xi = 60/(\pi \times 500^2)$, number of video files in the cloud is $F = 10^5$; size of the file requested from each UE is $x_f = 10^9$  bit, pathloss exponent is $\alpha = 5$, the SINR threshold is $T = 10$ dB, the probability that UE requests service from BS is $\eta = 0.014$ from  \cite{Anastasios2013}. Number of subbands $L = 6$. From (\ref{eqn:geometric_prog}), the delay threshold $D_{\text{th}} = 10^{-3}$ and $\gamma = 0.1$. We assume that there is a single server in cloud where the mean arrival time $\phi = 0.8$, mean service time $\tau = 5 \times 10^{-3}$, the coefficient of variation of inter-arrival time and service time are $c_a = 2$ and $c_s = 1$ as such, $\Ebb[D_{\text{bh}}] = 0.0051$ sec.

\begin{figure}[h]
\centering
\includegraphics[height=2.7 in, width=2.7 in, keepaspectratio = true]{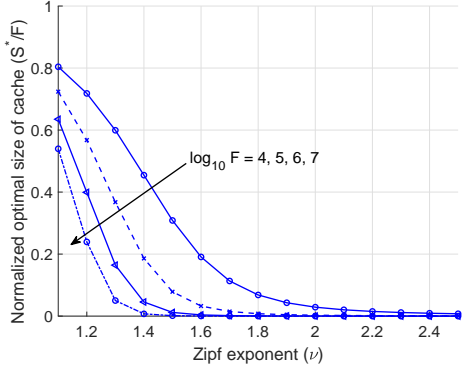}
\caption{Normalized optimal size of cache ($\frac{S^*}{F}$) versus $\nu$ when $\lambda = 20/(\pi\times 500^2)$}
\label{fig:OptS_nu}
\end{figure}
In Fig.~\ref{fig:OptS_nu}, the normalized optimal size of cache ($S^*/F$) versus Zipf exponent, $\nu$, with varying number of files in cloud ($F$)  from \tbf{Proposition \ref{prop:optimal-cache}} is shown. The bandwidth is $W = 300\times 10^6$. We assume the file in cloud is $F = 10^4, 10^5, 10^6$ and $10^7$. It can be seen that when $F$ increases, the $S^*/F$ decreases for given $\nu$. Also, with an increasing of $\nu$, the number of files to be stored in cache become less since greater $\nu$ means more skewness of file popularity.

\begin{figure}[h]
\centering
\includegraphics[height=2.7 in, width=2.7 in, keepaspectratio = true]{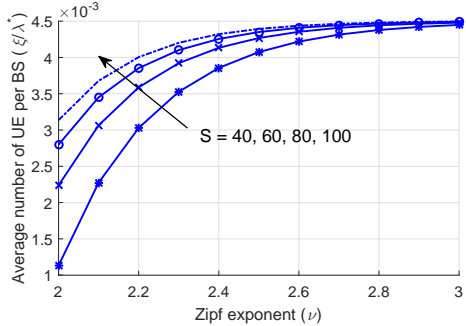}
\caption{Optimal number of UEs per BS ($\xi/\lambda^*$) versus $\nu$}
\label{fig:OptUEperBS_optlambda}
\end{figure}

\begin{figure}[h]
\centering
\includegraphics[height=2.7 in, width=2.7 in, keepaspectratio = true]{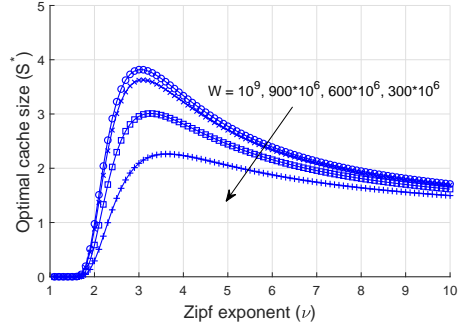}
\caption{Optimal size of cache ($S^*$) versus $\nu$}
\label{fig:OptSgeo}
\end{figure}

The optimal average number of UE per BS, $\xi/\lambda^*$, versus Zipf exponent, $\nu$ while varying the cache size, $S$ from \tbf{Proposition \ref{prop:optimal-lambda}} is plotted in Fig.~\ref{fig:OptUEperBS_optlambda} with $W = 300\times 10^6$. We see that when $\nu$ is increased $\xi/\lambda^*$ is also enhanced. This is because when increasing $\nu$, the number of files to be stored in the cache becomes smaller. Therefore, while $S^*$ decreases with an increasing of $\nu$, the BS intensity $\lambda^*$ increases. As such, $\xi/\lambda^*$ also enhances. For given $\nu$, when $S$ increases, $\xi/\lambda^*$ becomes higher. The optimal  cache size, $S^*$, versus $\nu$, from \tbf{Proposition \ref{prop:optSoptlambda}} is illustrated in Fig.~\ref{fig:OptSgeo}.


In Fig.~\ref{fig:OptSgeo}, the optimal cache size $S^*$ decreases when $W$ is decreased for a given $\nu$. This is because a higher bandwidth yields a greater throughput, , from $G \propto W$ in (\ref{eqn:goodput}), and since the constant $Q \propto C_2 \propto 1/G$, we have $q^* \propto Q \propto 1/G$. Thus, following \tbf{Proposition \ref{prop:optSoptlambda}}, we have $S^* \propto (1/q^*)^{1/(\nu-1)} \propto G^{1/(\nu-1)} \propto W^{1/(\nu-1)}$. Hence, for $\nu>1$, increasing $W$ leads to a decrease in the cost $q^*$, which in turn leads to an increased cache size $S^*$. On the other hand, since $R \propto 1/G$, the $G$ term cancels out in the expression for $\lambda^*$, making $\lambda^*$ independent of $W$. For increasing $\nu$, $S^*$ decreases as such, $\lambda^*$ increases.

\section{Conclusion}
\label{subsec:conclusion}
 We have modeled and analyzed the performance of large scale cache-enabed cellular network. We have obtained the downlink SINR coverage probability and thoughput based on stochastic geometry analysis. Based on these results with the proactive caching placement, we have analyzed the fronthaul and backhaul delay. The minimization problem through the reduction of the cache size and/or BS intensity constrain on the probabilistic delay is proposed. We have solved the optimization problem when either BS intensity or the cache size is held fixed and also when both BS intensity and cache size are vary. Then, the optimization problem when both BS intensity and cache size are variable subject to latency constraint is converted into a geometric program and the closed-form solution is obtained. We have observed that the Zipf exponent has significant effect on performance of MNO.

\section*{Acknowledgment}
This work has been financially supported by 6Genesis (6G) Flagship project (grant 318927). Also, the authors would like to thank Professor Ekram Hossain, University of Manitoba, Canada and Professor Dusit Niyato,  Nanyang Technological University, Singapore for their helpful comments and suggestions.


\appendix
The generalized harmonic number $H_{S,\nu}$ does not have a closed form expression. Nevertheless, for analytical tractability, we can make an asymptotic approximation\footnote{Here $f(x) \sim g(x)$ if and only if $\lim_{x\to \infty}\frac{f(x)}{g(x)} = 1$.} in terms of $S$ and $\nu$. To do so, we will relate the generalized harmonic number to the Hurwitz zeta function and then use the properties of Hurwitz zeta function. The Hurwitz zeta function, $\zeta(s,a)$, is defined as \cite[Eqn 25.11.1]{DLMF2017}
\beq
\zeta(s,a) = \sum_{n=0}^{\infty} \frac{1}{(n+a)^s},
\label{eqn:hurwitz-zeta-fun}
\eeq
 where $\Re(s) >1$ and $a \neq 0, -1, -2, \ldots$. The Hurwitz zeta function reduces to the Reimann zeta function when $a=1$,
$ \zeta(s,1) =\zeta(s)$, where $\zeta(s)$ is the Riemann zeta function. Also, harmonic sums can be expressed in terms of Hurwitz zeta function as \cite[Eqn 25.11.4]{DLMF2017}
\beq
\sum_{n=0}^{m-1} \frac{1}{(n+a)^{s}} = \zeta(s,a) - \zeta(s, a+m).
\label{eqn:hurwitz-zeta-harmonic-sum-1}
\eeq

For our case, comparing (\ref{eqn:def-genearlized-harmonic-sum}) and (\ref{eqn:hurwitz-zeta-harmonic-sum-1}), we can express the generalized harmonic sum $H_{S,\nu}$ in terms of the Hurwitz zeta function as
\beq
H_{S,\nu} = \sum_{n=0}^{S-1}\frac{1}{(n+1)^\nu} = \zeta(\nu) - \zeta(\nu,S+1)
\label{eqn:hurwitz-zeta-harmonic-sum-2}
\eeq

Now, as $S \to \infty$, the asymptotic expansion of Hurwitz zeta function is given by \cite[Eqn 25.11.43]{DLMF2017}
\begin{align}
\zeta(\nu,S+1) \sim & \frac{(S+1)^{1-\nu}}{\nu-1} + \frac{1}{2}(S+1)^{-\nu} \nonumber \\ & \quad + \sum_{k=1}^{\infty} \frac{B_{2k}}{(2k)!}(\nu)_{2k-1}(s+1)^{1-\nu-2k},
\label{eqn:hurwitz-zeta-asymptotic}
\end{align}
where $B_{2k}$ are Bernoulli numbers and $(\nu)_{2k-1} = \nu (\nu + 1) \cdots (\nu + 2k - 2)$ are Pochammer's symbol for rising factorial.

Taking only the first dominant term from (\ref{eqn:hurwitz-zeta-asymptotic}) and substituting it in (\ref{eqn:hurwitz-zeta-harmonic-sum-2}), we obtain the asymptotic approximation for the generalized harmonic number as
\beq
H_{S,\nu} \sim \zeta(\nu) - \frac{(S+1)^{1-\nu}}{\nu -1}.
\label{eqn:harmonic-sum-asymptotic}
\eeq

Thus, from the above arguments, using (\ref{eqn:hit-probability}) and (\ref{eqn:harmonic-sum-asymptotic}), we have desired lemma.

\bibliographystyle{IEEE}

\end{document}